\documentclass[11pt]{article}

\def\ie{\emph{i.e.}}

\usepackage{latexsym}
\usepackage[usenames,dvipsnames]{color}
\usepackage{verbatim}
\usepackage{axodraw}
 \usepackage[english]{babel}
 \usepackage{graphicx}				
 \usepackage{geometry}
 \usepackage{amssymb}
 \usepackage{pict2e}
 \usepackage{amsthm}
 \usepackage{booktabs}
\usepackage{authblk}
\title{A shoving model for collectivity in hadronic collisions}
\author[1]{Christian Bierlich}
\author[2]{G{\"o}sta Gustafson}
\author[3]{Leif L{\"o}nnblad}
\affil[1]{Lund University, Email: \texttt{christian.bierlich@thep.lu.se} }
\affil[2]{Lund University, Email: \texttt{gosta.gustafson@thep.lu.se} }
\affil[3]{Lund University, Email: \texttt{leif.lonnblad@thep.lu.se} }
\begin{document}
\begin{flushright}
  MCnet-16-48\\
  LU-TP 16-64\\
  arXiv:1612.nnnn [hep-ph]
\end{flushright}
\vspace*{1cm}
\begin{center}
  \LARGE A shoving model for collectivity in hadronic collisions \normalsize\\[8mm]
  Christian Bierlich, G{\"o}sta Gustafson, Leif L{\"o}nnblad\\[5mm]
  Department of Astronomy and Theoretical Physics, Lund
  University\footnote{Email: \texttt{christian.bierlich@thep.lu.se},
    \texttt{gosta.gustafson@thep.lu.se},
    \texttt{leif.lonnblad@thep.lu.se}}$^{,}$\footnote{Work supported in part by the MCnetITN FP7 Marie Curie Initial Training Network, contract PITN-GA-2012-315877, and the Swedish Research Council (contracts 621-2012-2283 and 621-2013-4287).}
\end{center}
\vspace*{1cm}
\textsc{Abstract:} An extension of the rope hadronization
model, which has previously provided good descriptions of
hadrochemistry in high multiplicity pp collisions, is presented. The extension includes a
dynamically generated transverse pressure, produced by the excess
energy from overlapping strings. We find that this model can
qualitatively reproduce soft features of Quark Gluon Plasma in small
systems, such as higher $\langle p_\perp \rangle$ for heavier particles and
long range azimuthal correlations forming a ridge. The effects are similar
to those obtained from a hydrodynamic expansion, but without assuming a
thermalized medium. 

\vspace*{0.6cm}
Recent precise measurements of pp and p$A$ collisions at the LHC show
flow-like effects
\cite{Aaboud:2016yar,Khachatryan:2016txc,Abelev:2012ola} similar to
those found in high energy nucleus collisions. Examples are ridge like
structures, quantified in different flow coefficients, and
measurements of strangeness enhancement with increasing event
activity \cite{Adam:2016emw}. These are regarded as two important
characteristics of the soft features of the Quark Gluon Plasma, and are
often described in a hydrodynamical framework assuming
thermal equilibrium.
 
Dynamical models based on string
\cite{Andersson:1983jt,Andersson:1983ia} or cluster
\cite{Marchesini:1983bm} hadronization models, \emph{e.g.}\
\textsc{Pythia8} \cite{Sjostrand:2006za,Sjostrand:2014zea} and HERWIG7
\cite{Bellm:2015jjp}, are able to describe the general soft features
of pp collisions in a very satisfactory way. The need for imposing new
dynamics at a macroscopic level, only present at soft scales, is
complicated, as it breaks the principle of jet universality by
introducing a scale below which new dynamics should be "switched
on". A quite successful model based on this principle, is the
core--corona model \cite{Werner:2007bf}, implemented in the EPOS
generator \cite{Pierog:2013ria}, where events are subdivided into
"core" and "corona" events based on event activity. Recently, attempts
has been made to incorporate a ``thermal'' exponential
$m_\perp$-spectrum for the string break-ups in the Lund hadronization
model \cite{Fischer:2016zzs} with promising results, but it is still
unclear whether such a "microscopization" can capture the essential
features of hydrodynamics, and if so, if this picture can correctly
describe both hadrochemistry and flow.

To provide a description of the hadrochemistry in the underlying event of pp
collisions, we recently suggested a "rope hadronization" model
\cite{Bierlich:2014xba}, based on work by Biro, Knoll and Nielsen
\cite{Biro:1984cf}. This model provides corrections to the string
hadronization model, by allowing strings overlapping in transverse space to
act coherently as a "rope". The model is implemented in the DIPSY event
generator \cite{Flensburg:2011kk}, which provides a dynamical picture of the
event structure in impact parameter space,
allowing for a calculation of the colour field strength in each small rope
segment\footnote{A Lund string is in its simplest form, a straight piece
stretched between a quark and an anti-quark, or a colour triplet and
anti-triplet. As gluons are added to the 
string, they act as point-like "kinks" on the string, carrying energy and
momentum \cite{Andersson:1979ij}. We will denote all straight pieces between gluons or
(anti)quarks string segments. A $q-g-\bar{q}$ string thus has two segments.}. 
This formalism also includes all
fluctuations. The colour field is characterized by two 
quantum numbers $\{p,q\}$, which together signifies its SU(3) multiplet
structure. Lattice calculations have shown \cite{Bali:2000un}, that the string
tension -- energy per unit length -- scales with the quadratic Casimir
operator of the multiplet, such that the ratio of the enhanced rope tension
($\tilde{\kappa}$) to the triplet string tension in vacuum ($\kappa$) is: 
\begin{equation}
	\frac{\tilde{\kappa}}{\kappa} = \frac{1}{4}\left(p^2 + q^2 + pq +
        3(p+q) \right). 
\end{equation}
The enhancement of string tension was shown \cite{Bierlich:2015rha} to greatly
influence the ratio of strange to non--strange hadrons, and to give the correct
dependence on event activity as measured by ALICE \cite{Adam:2016emw}.

\begin{figure}
\begin{picture}(300,100)(0,0)
	\Oval(60,50)(7,7)(0)
	\Oval(64,50)(7,7)(0)
	\Oval(56,50)(7,7)(0)
	\Oval(60,46)(7,7)(0)
	\Oval(60,54)(7,7)(0)
	\Text(60,0)[c]{$t = t_1$}
	\Oval(150,50)(7,7)(0)
	\Oval(158,50)(7,7)(0)
	\Oval(142,50)(7,7)(0)
	\Oval(150,42)(7,7)(0)
	\Oval(150,58)(7,7)(0)
	\LongArrow(158,50)(168,50)
	\LongArrow(142,50)(132,50)
	\LongArrow(150,58)(150,68)
	\LongArrow(150,42)(150,32)
	\Text(150,0)[c]{$t = t_2$}
	\Oval(240,50)(7,7)(0)
	\Oval(256,50)(7,7)(0)
	\Oval(224,50)(7,7)(0)
	\Oval(240,34)(7,7)(0)
	\Oval(240,66)(7,7)(0)
	\LongArrow(256,50)(271,50)
	\LongArrow(224,50)(209,50)
	\LongArrow(240,34)(240,19)
	\LongArrow(240,66)(240,81)
	\Text(240,0)[c]{$t = t_3$}
	\Oval(330,50)(7,7)(0)
	\Oval(355,50)(7,7)(0)
	\Oval(305,50)(7,7)(0)
	\Oval(330,25)(7,7)(0)
	\Oval(330,75)(7,7)(0)
	\LongArrow(355,50)(370,50)
	\LongArrow(305,50)(290,50)
	\LongArrow(330,25)(330,10)
	\LongArrow(330,75)(330,90)
	\Text(330,0)[c]{$t = t_4$}
\LongArrow(5,5)(20,5)
\LongArrow(5,5)(5,20)
\Text(-5,20)[c]{$b_y$}
\Text(20,-2)[c]{$b_x$}
\end{picture}
\caption{\label{fig:stringshoveCartoon}Cartoon in impact parameter space
  showing strings overlapping at time $t=t_1$, and as time progresses ($t_1 <
  t_2 < t_3 < t_4$), they move apart, picking up $p_\perp$ as indicated with
  arrows.} 
\end{figure}
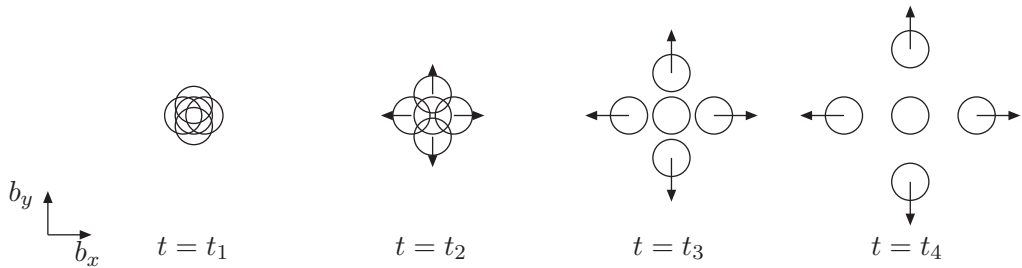

In this letter we show how this enhanced string tension can also be employed
to generate a flow--like effect. Since the energy density in the overlap
region is higher than outside, a pressure will be dynamically generated,
pushing strings outwards. In figure~\ref{fig:stringshoveCartoon} this
principle is 
illustrated; we sketch several overlapping strings in impact parameter space
at some initial time $t_1$. The density is larger towards center, giving a
pressure gradient. We start a spatio-temporal evolution and let the strings
pick up transverse momentum from the excess energy in the overlap regions. As
the strings move further apart, the excess energy will decrease, and so will
the transverse pressure. From time $t_1$ to $t_2$ the strings pick up
some transverse momentum, as indicated with arrows, and move a little
bit. From $t_2$ to $t_3$ the strings move, but picks up less
transverse momentum, as the overlap is now smaller. From $t_3$ to $t_4$ 
the strings only move, and picks up no transverse momentum, as there is
no overlap. The strings should of course hadronize at some point along the
way, and we interrupt the evolution at some given time, where strings are no
longer allowed to pick up $p_\perp$ or propagate. 

The partonic state obtained from the DIPSY MC is formulated in
rapidity ($y$) and transverse coordinate space
($\mathbf{b}_\perp$). Colour-connected partons separated by a distance
$\Delta \mathbf{b}_\perp$ are also given opposite transverse momenta
$\mathbf{p}_\perp \approx \Delta \mathbf{b}_\perp/(\Delta
\mathbf{b}_\perp)^2$.  The initial state is two Lorentz contracted
pancakes colliding at $z=0$, and the string segments are then
stretched out mainly along the $z$ direction.  The distribution of
gluons is approximately boost invariant, and to visualize the effect
of the transverse repulsion, it is most easy to study a string segment
stretched between two gluons in a system where they have rapidities
$\pm \Delta y/2$. The endpoints of this string segment will then move
out with longitudinal velocities $v_L=\pm \tanh(\Delta y/2)$, and the
length of the segment in coordinate space, at time $t$, is
consequently $ t\!\cdot\! \tanh(\Delta y)$. The repulsive transverse
force between two strings is proportional to the length of the
overlapping region, and is therefore proportional to $f\!\cdot\!
t\cdot\! \Delta y$, where $f$ is the force per unit string length.

The cartoon in figure~\ref{fig:stringshoveCartoon} represents in a
schematic way a "slice" in rapidity\footnote{In reality the strings
  are, of course, not distributed symmetrically, instead there are
  large fluctuations in the transverse positions of the strings.}.
The result of the repulsion will be a transverse velocity for the
string, which might be represented by very many very soft gluons. The
breakup of such a string state cannot be handled current implementations 
of string hadronization, as in \emph{e.g.}\ \textsc{Pythia8}. As the DIPSY 
generator interfaces to the \textsc{Pythia8} hadronization implementation,
this must be remedied.
A transverse gluon will give momentum to hadrons within one unit of rapidity on either
side of the gluon. It is therefore possible to simulate the effect of
the continuous distribution of infinitely soft gluons by finite gluons
separated by at most one rapidity unit.  In our calculations we cut
the event into many rapidity slices, and in each slice we let the
strings ``shove'' each other apart. The mechanism for shoving is to
add a small excitation (\ie~a gluon) to each string in each slice. In
each time--step $\delta t$ a string within a slice $\delta y$ (and
thus length $\delta l= t\, \delta y$) will get a kick in the
transverse direction $\delta p_\perp = f\, t\, \delta y \, \delta t$.
As the mass of the string piece is $\approx \kappa \,\delta l = \kappa\, t\, \delta y$
also is proportional to the time $t$, we note that the factors $t$
drop out in the result for the transverse velocity boost. When the
strings no longer overlap, the many small kicks are added to a set of
gluons, which can be handled by \textsc{Pythia8}.
The $p_\perp$ of these gluons are chosen sufficiently small, so that
they have lost their energy before the string hadronizes. This implies
that their transverse momenta do not produce a jet, but just some
extra $p_\perp$ within a rapidity range $\pm 1$ unit. The result is
then not sensitive to the exact number of gluons within such an
interval, as long as their transverse $p_\perp$ add up to the same
value.

In our current implementation, the shoving is implemented as the sum
of many small kicks between all pairs of string segments in
different rapidity intervals spread out evenly in the available phase
space. This is done in several time steps, and in each such kick, the
momentum is conserved as the inserted gluons will get equal and
opposite transverse kicks, while the longitudinal recoils are absorbed
by the original partons in the end of the string segments.

%
%

If the string is similar to a flux tube in a type I superconductor,
the field is approximately constant within a cylindrical tube. Such a
picture was used in analytic studies by Abramovsky and coworkers as
early as 1988 \cite{Abramovsky:1988zh}. At that time there was no
experimental evidence for long range azimuthal correlations, but the
model was recently revived in ref.~\cite{Altsybeev:2015vma}, and
implemented as a Monte Carlo toy model. In this model the increased energy per unit string length, $\frac{\delta E}{\delta l}$, scales with the overlap area in
transverse space. This gives (in our notation):
\begin{equation}
	\frac{\delta E}{\delta l} = \Theta(R - d)\sqrt{\left(\kappa +
	\tilde{\kappa}\frac{A(R,d_\perp)}{\pi R^2} \right)^2 - \kappa^2}, 
\end{equation}
where $R$ is a characteristic transverse radius of the cylinder, and
$A(R,d_\perp)$ is the overlap area between two circles of radius $R$,
sitting at a distance $d_\perp$ apart in the transverse plane. Due to the repulsion, this energy is transferred to kinetic energy, giving a transverse velocity boost to the strings.

For a type II superconductor the
field strength falls off more smoothly away from a central
core. Lattice calculations favour a QCD string with properties on the
border between type I and II \cite{Cea:2014uja}. In our implementation
we have assumed a smooth Gaussian form, similar to the lattice result, and furthermore added a
temporal part to the evolution, as described above:
\begin{equation}
	\frac{dp_\perp}{dt\, dl} = \frac{g\kappa}{R \sqrt{2
	\pi}}\exp\left(-\frac{d^2_\perp(t)}{2R^2} \right).
\end{equation}
Here $g$ is a free parameter controlling the strength of the shoving,
which should be of order unity. The transverse distance between the strings has acquired explicit time dependence. 
As discussed above, such a repulsion gives a transverse boost to the
string segments, and thus extra $p_\perp$ to heavier hadrons. The
$\langle p_\perp\rangle$ dependence on the hadronic mass, is an
observable which is often connected to hydrodynamics, as a
thermodynamic pressure would provide the same physical effect.

\begin{figure}
	\centering
	\includegraphics[width=0.8\textwidth]{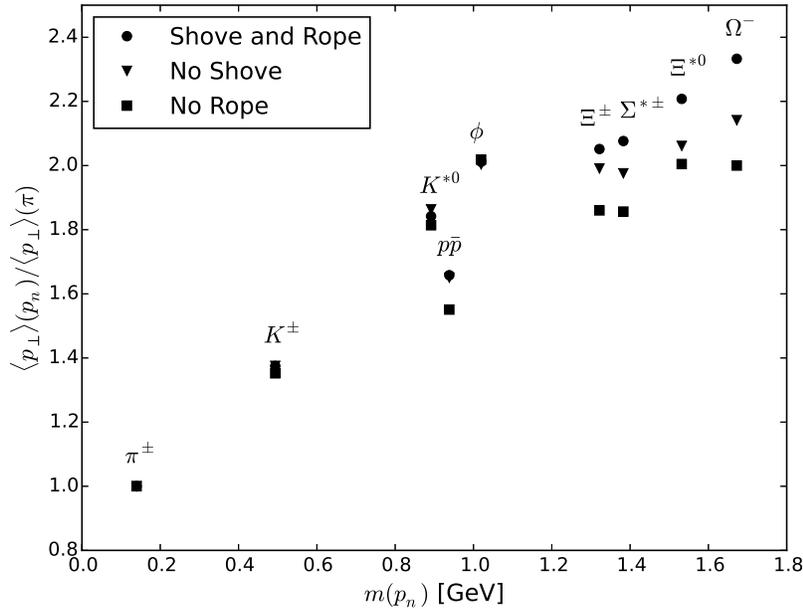}
	\caption{\label{fig:meanpt} Average $p_\perp$ as a function of
          hadronic mass, for several species. Results are presented
          for DIPSY without ropes, DIPSY with ropes and DIPSY with
          ropes and shoving.}
\end{figure}

In figure~\ref{fig:meanpt} we show the $\langle p_\perp\rangle$ for
several hadron species, divided by $\langle p_\perp\rangle$ for pions,
in pp collisions at $\sqrt{s}=7$~TeV. We show results for DIPSY without
ropes, with ropes but no shoving, and with both ropes and shoving. By
choosing a ratio, rather than the raw $\langle p_\perp \rangle$, we
minimize effects from small differences in the tuning of the three
models. Even DIPSY without ropes shows a rise. This is expected, as
lighter hadrons are more likely to be decay products -- consequently
with lower $\langle p_\perp \rangle$ -- than heavy particles
originating directly from the string breaking. When ropes are switched
on, the $\langle p_\perp \rangle$ rises slightly. This is an effect of
the enhanced string tension. When the string breaks, the emerging
hadron obtains a $p_\perp$ taken from a Gaussian distribution. The
width of this distribution rises with the effective string tension as
\cite{Bierlich:2014xba}:
\begin{equation}
	\tilde{\sigma}_\perp = \sigma_\perp\sqrt{\frac{\tilde{\kappa}}{\kappa}}.
\end{equation}

The rise in $\langle p_\perp\rangle$ from ropes is, however, not
directly mass dependant, while the expected mass dependence in the
effect from shoving is clearly seen in figure~\ref{fig:meanpt}.

Di-hadron correlations, which in data show a ridge effect, are of
particular interest. We know that the DIPSY generator has problems
reproducing the high-$p_\perp$ end of charged particle spectra; it
generates too many hard partons. This introduces a potential problem
for our string shoving model, which assumes parallel strings. To study
the correlation effects on pairs of soft hadrons, we have therefore
biased the generated events (pp at $\sqrt{s}=7$~TeV) by only
considering strings that span a rapidity range larger than $\Delta y =
8$, and with no partons above $p_\perp=3$~GeV. Thus we get events with
long strings, almost parallel in rapidity.\footnote{Note that the
  results on the mass dependence of $\langle p_\perp\rangle$ are
  fairly insensitive to this bias, and in figure~\ref{fig:meanpt} the
  bias was not applied.}

\begin{figure}
	\centering
	\includegraphics[width=0.45\textwidth]{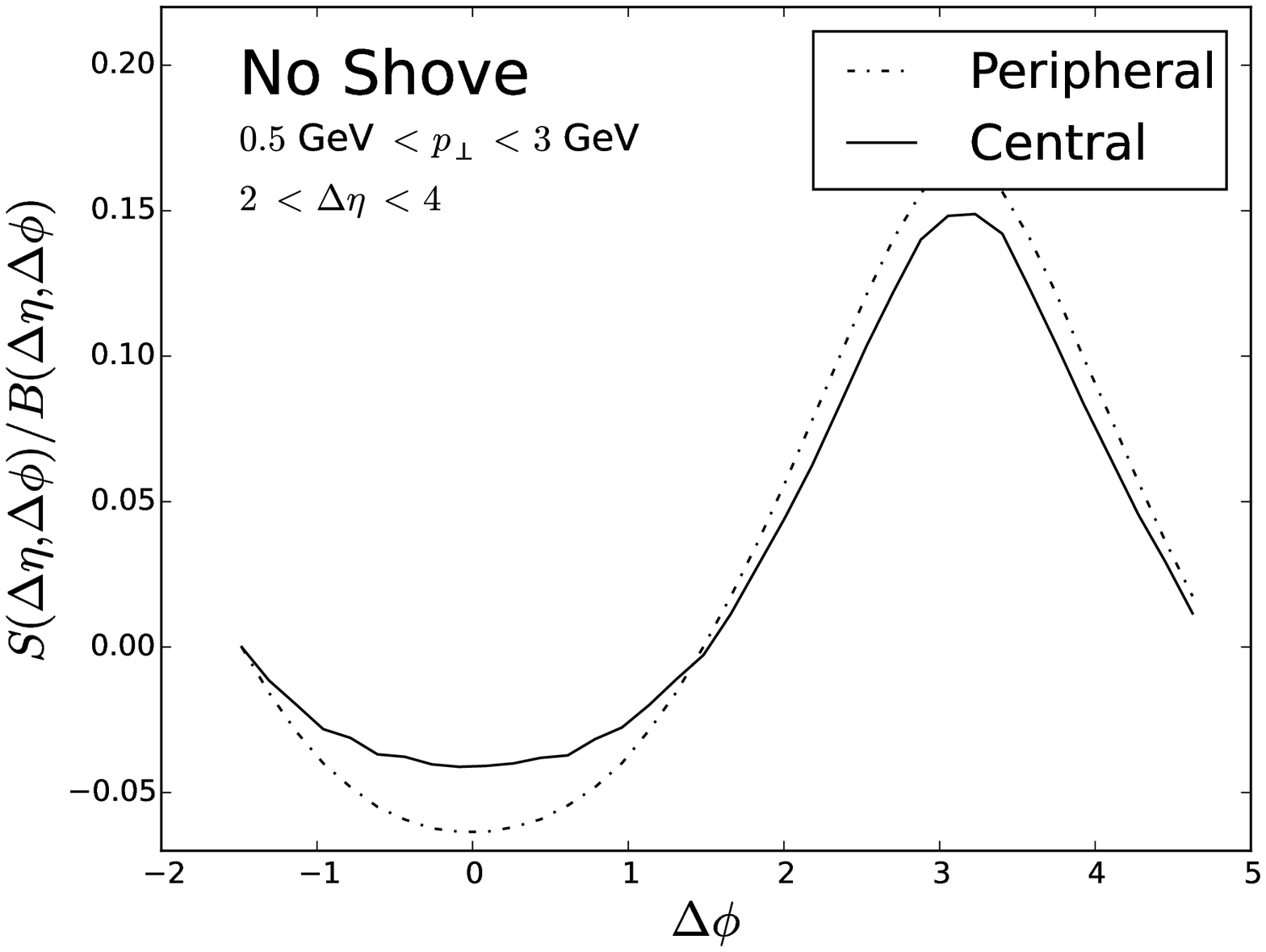}
	\includegraphics[width=0.45\textwidth]{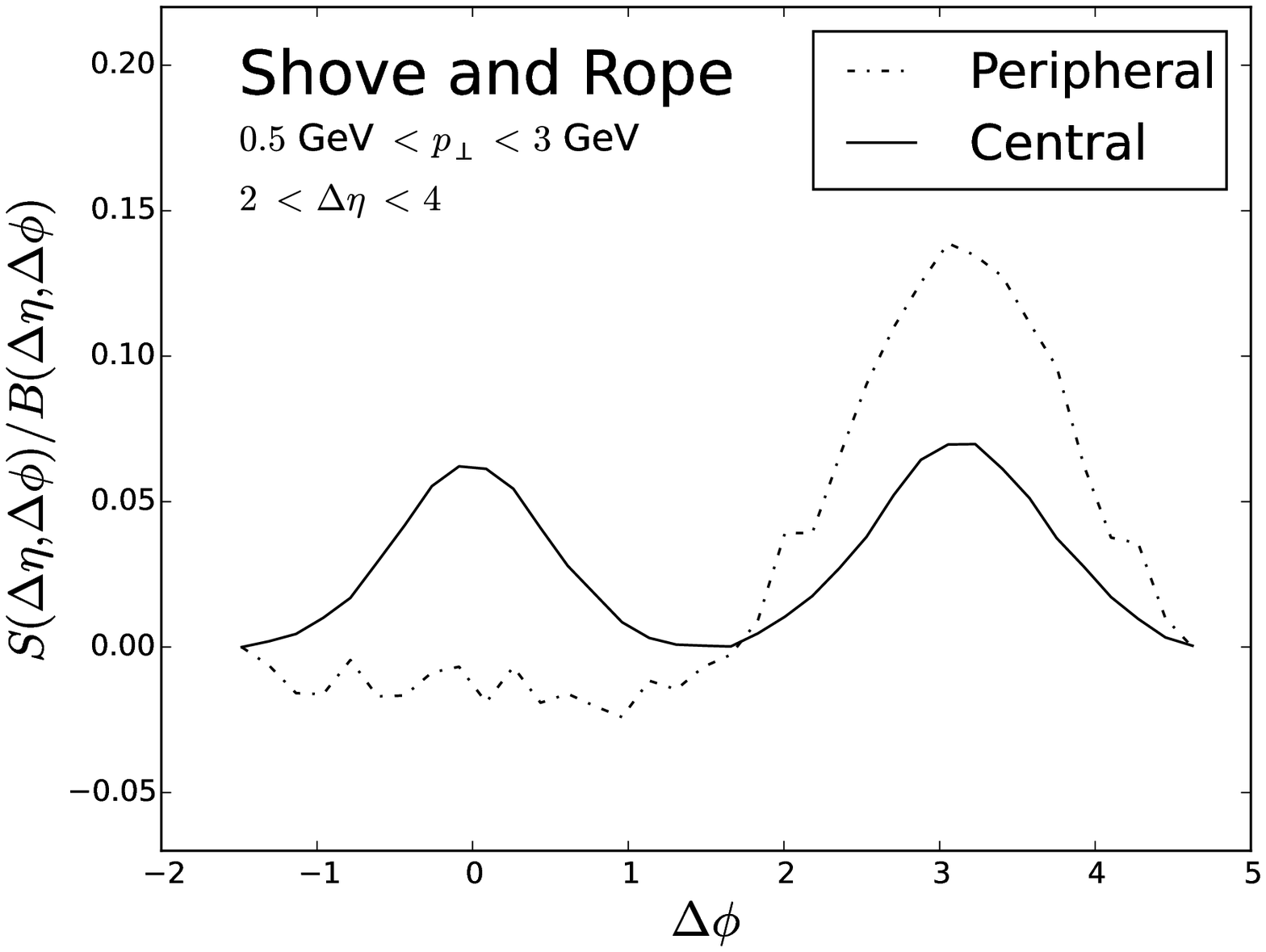}
	\caption{\label{fig:correlations} Two particle correlations
          without (left) and with (right) shoving effects, for central
          and peripheral pp events at $\sqrt{s} = 7$~TeV.}
\end{figure}

To calculate the correlations, we employ an analysis similar to the
one chosen by experiments \cite{Aaboud:2016yar,Khachatryan:2016txc},
where a signal distribution $S(\Delta \phi, \Delta \eta)$ is divided
by a random background distribution, $B(\Delta \phi, \Delta \eta)$, constructed by combining
particles from two different events in the same centrality class. In
figure~\ref{fig:correlations} we show results for $2<\Delta\eta<4$,
for particles with transverse momentum between $0.5$ and $3$~GeV,
using the rope model in DIPSY without (left) and with (right) shoving
effects. We see the emergence of a clear "ridge" around
$\Delta\eta=0$. Although the emerging ridge is roughly of the same
relative size as the one recently measured by ATLAS
\cite{Aaboud:2016yar} in high multiplicity pp events at
$\sqrt{s}=13$~TeV, we stress that the results presented here can only
be taken as a qualitative proof-of-concept, due to the event bias we
have introduced.

In this letter we have demonstrated that by introducing shoving in the
rope hadronization mechanism implemented in DIPSY, we are able to
qualitatively describe collective phenomena in pp collisions, in
addition to the quantitatively correct description of hadrochemistry
already provided by the ropes. We remark that the mechanism does not
require any medium or thermalization, but is composed solely of
microscopic interactions.

To get a better quantitative description of collective phenomena,
further studies are needed. High $p_\perp$ jets are expected to rapidly leave
the dense system of strings, and more work is needed for a realistic
description of the interplay beteen the jet and the rope.
 Also, as it stands, the model introduces a number of
parameters, \emph{e.g.}, the strength of the shoving, the number of
rapidity intervals, and the number and size of the time
steps. Although all of them have physically motivated values, their
influence on the results must be studied in more detail and, in the
end, they need to be tuned to data.

Finally we note that, if successfully tuned to pp data, the model can
be directly applied to collisions involving heavy ions in DIPSY,
and in that way provide a complementary picture to the conventional
hydrodynamical description of p$A$ and $AA$ collisions. In addition we
plan to implement the model to our developing heavy ion event
generator based on \textsc{Pythia8} presented in ref.~\cite{Bierlich:2016smv}.

\bibliographystyle{ieeetr}  
\bibliography{shoving} 
\end{document}